\documentclass[runningheads]{llncs}
\usepackage[T1]{fontenc}

\usepackage{subcaption}
\usepackage{verbatim}
\usepackage{color}

\usepackage{isabelle}
\usepackage{isabellesym}
\usepackage{listings}

\usepackage{hyperref}
\usepackage{prettyref}

\usepackage{amsmath}
\usepackage{amssymb}
\usepackage{tikz}
\usepackage[normalem]{ulem}
\usepackage{xspace}
\usepackage{mathtools}

\usetikzlibrary{calc}
\usetikzlibrary{patterns}
\usetikzlibrary{decorations.pathreplacing}
\definecolor{darkgreen}{HTML}{008400}

\newcommand{\pickexample}[1]{
\begin{tikzpicture}[scale=#1]
    \foreach \i in {0,...,5} {
        \foreach \j in {0,...,3} {
            \node[circle, fill, inner sep=1.5pt] at (\i, \j) {};
        }
    }

    \coordinate (A) at (1, 0);
    \coordinate (B) at (2, 1);
    \coordinate (C) at (4, 0);
    \coordinate (D) at (5, 2);
    \coordinate (E) at (4, 2);
    \coordinate (F) at (3, 2);
    \coordinate (G) at (2, 3);
    \coordinate (H) at (1, 3);
    \coordinate (I) at (0, 3);
    \draw [thick] (A) -- (B) -- (C) -- (D) -- (E) -- (F) -- (G) -- (H) -- (I) -- cycle;

    \node[circle, fill=blue, inner sep=1.5pt] at (A) {};
    \node[circle, fill=blue, inner sep=1.5pt] at (B) {};
    \node[circle, fill=blue, inner sep=1.5pt] at (C) {};
    \node[circle, fill=blue, inner sep=1.5pt] at (D) {};
    \node[circle, fill=blue, inner sep=1.5pt] at (E) {};
    \node[circle, fill=blue, inner sep=1.5pt] at (F) {};
    \node[circle, fill=blue, inner sep=1.5pt] at (G) {};
    \node[circle, fill=blue, inner sep=1.5pt] at (H) {};
    \node[circle, fill=blue, inner sep=1.5pt] at (I) {};

    \node[circle, fill=darkgreen, inner sep=1.5pt] at (1, 1) {};
    \node[circle, fill=darkgreen, inner sep=1.5pt] at (1, 2) {};
    \node[circle, fill=darkgreen, inner sep=1.5pt] at (2, 2) {};
    \node[circle, fill=darkgreen, inner sep=1.5pt] at (3, 1) {};
    \node[circle, fill=darkgreen, inner sep=1.5pt] at (4, 1) {};
\end{tikzpicture}
}

\newcommand{\goodlinepath}[1]{
    \begin{tikzpicture}[scale=#1]
        \foreach \i in {0,...,5} {
            \foreach \j in {0,...,3} {
                \node[circle, fill, inner sep=1.5pt] at (\i, \j) {};
            }
        }
    
        \coordinate (A) at (1, 0);
        \coordinate (B) at (2, 1);
        \coordinate (C) at (4, 0);
        \coordinate (D) at (5, 2);
        \coordinate (E) at (4, 2);
        \coordinate (F) at (3, 2);
        \coordinate (G) at (2, 3);
        \coordinate (H) at (1, 3);
        \coordinate (I) at (0, 3);
        \fill [blue!20] (A) -- (B) -- (C) -- (H) -- (I) -- (A);
        \fill [pink!50] (C) -- (D) -- (E) -- (F) -- (G) -- (H) -- (C);
        \draw [thick] (A) -- (B) -- (C) -- (D) -- (E) -- (F) -- (G) -- (H) -- (I) -- cycle;
        \draw [thick, darkgreen] (C) -- (H);

        \node[circle, fill=black, inner sep=1.5pt] at (A) {};
        \node[circle, fill=black, inner sep=1.5pt] at (B) {};
        \node[circle, fill=black, inner sep=1.5pt] at (C) {};
        \node[circle, fill=black, inner sep=1.5pt] at (D) {};
        \node[circle, fill=black, inner sep=1.5pt] at (E) {};
        \node[circle, fill=black, inner sep=1.5pt] at (F) {};
        \node[circle, fill=black, inner sep=1.5pt] at (G) {};
        \node[circle, fill=black, inner sep=1.5pt] at (H) {};
        \node[circle, fill=black, inner sep=1.5pt] at (I) {};
    
        \node[circle, fill=black, inner sep=1.5pt] at (1, 1) {};
        \node[circle, fill=black, inner sep=1.5pt] at (1, 2) {};
        \node[circle, fill=black, inner sep=1.5pt] at (2, 2) {};
        \node[circle, fill=black, inner sep=1.5pt] at (3, 1) {};
        \node[circle, fill=black, inner sep=1.5pt] at (4, 1) {};
    \end{tikzpicture}
}

\newcommand{\goodpolygonalpath}[1]{
    \begin{tikzpicture}[scale=#1]
        \foreach \i in {0,...,5} {
            \foreach \j in {0,...,3} {
                \node[circle, fill, inner sep=1.5pt] at (\i, \j) {};
            }
        }
    
        \coordinate (A) at (1, 0);
        \coordinate (B) at (2, 1);
        \coordinate (C) at (4, 0);
        \coordinate (D) at (5, 2);
        \coordinate (E) at (4, 2);
        \coordinate (F) at (3, 2);
        \coordinate (G) at (2, 3);
        \coordinate (H) at (1, 3);
        \coordinate (I) at (0, 3);
        \fill [blue!20] (A) -- (B) -- (C) -- (4, 1) -- (2, 2) -- (1, 1) -- (H) -- (I) -- (A);
        \fill [pink!50] (C) -- (D) -- (E) -- (F) -- (G) -- (H) -- (1, 1) -- (2, 2) -- (4, 1) -- (C);
        \draw [thick] (A) -- (B) -- (C) -- (D) -- (E) -- (F) -- (G) -- (H) -- (I) -- cycle;
        \draw [thick, darkgreen] (C) -- (4, 1) -- (2, 2) -- (1, 1) -- (H);

        \node[circle, fill=black, inner sep=1.5pt] at (A) {};
        \node[circle, fill=black, inner sep=1.5pt] at (B) {};
        \node[circle, fill=black, inner sep=1.5pt] at (C) {};
        \node[circle, fill=black, inner sep=1.5pt] at (D) {};
        \node[circle, fill=black, inner sep=1.5pt] at (E) {};
        \node[circle, fill=black, inner sep=1.5pt] at (F) {};
        \node[circle, fill=black, inner sep=1.5pt] at (G) {};
        \node[circle, fill=black, inner sep=1.5pt] at (H) {};
        \node[circle, fill=black, inner sep=1.5pt] at (I) {};
    
        \node[circle, fill=black, inner sep=1.5pt] at (1, 1) {};
        \node[circle, fill=black, inner sep=1.5pt] at (1, 2) {};
        \node[circle, fill=black, inner sep=1.5pt] at (2, 2) {};
        \node[circle, fill=black, inner sep=1.5pt] at (3, 1) {};
        \node[circle, fill=black, inner sep=1.5pt] at (4, 1) {};
    \end{tikzpicture}
}

\newcommand{\trianglethreesplit}[1]{
\begin{tikzpicture}[scale=#1]
    \coordinate (A) at (0, 0);
    \coordinate (B) at (1, 2);
    \coordinate (C) at (2, 1);
    \coordinate (Z) at (1, 1);

    \draw[thick] (A) -- (Z);
    \draw[thick] (B) -- (Z);
    \draw[thick, fill=blue!20] (A) -- (B) -- (Z) -- (A);

    \draw (2.5, 1) edge[->] (3.5, 1);

    \foreach \i in {0,...,2} {
        \foreach \j in {0,...,2} {
            \node[circle, fill, inner sep=1.5pt] at (\i, \j) {};
        }
    }
    \node[circle, fill=darkgreen, inner sep = 1.5pt] at (Z) {};

    \coordinate (A) at (4, 0);
    \coordinate (B) at (5, 2);
    \coordinate (C) at (6, 1);
    \coordinate (Z) at (5, 1);

    \draw[line width=0pt, fill=orange!20] (A) -- (Z) -- (B) -- (A);
    \draw[line width=0pt, fill=blue!20] (C) -- (Z) -- (B) -- (C);

    \draw[thick] (A) -- (Z);
    \draw[thick, darkgreen] (B) -- (Z);
    \draw[thick] (A) -- (B) -- (C) -- (Z) -- (A);

    \draw (6.5, 1) edge[->] (7.5, 1);

    \foreach \i in {4,...,6} {
        \foreach \j in {0,...,2} {
            \node[circle, fill, inner sep=1.5pt] at (\i, \j) {};
        }
    }
    \node[circle, fill=darkgreen, inner sep = 1.5pt] at (Z) {};

    \coordinate (A) at (8, 0);
    \coordinate (B) at (9, 2);
    \coordinate (C) at (10, 1);
    \coordinate (Z) at (9, 1);

    \draw[line width=0pt, fill=blue!20] (A) -- (Z) -- (C) -- (A);
    \draw[line width=0pt, fill=orange!20] (A) -- (Z) -- (C) -- (B) -- (A);

    \draw[thick, darkgreen] (A) -- (Z);
    \draw[thick, darkgreen] (C) -- (Z);
    \draw[thick] (A) -- (B) -- (C) -- (A);

    \foreach \i in {8,...,10} {
        \foreach \j in {0,...,2} {
            \node[circle, fill, inner sep=1.5pt] at (\i, \j) {};
        }
    }
    \node[circle, fill=darkgreen, inner sep = 1.5pt] at (Z) {};
\end{tikzpicture}
}

\newcommand{\goodlinepathconvex}[1]{
\begin{tikzpicture}[scale=#1]
    \coordinate (A) at (0, 0);
    \coordinate (B) at (1, -0.25);
    \coordinate (C) at (2.25, 0.35);
    \coordinate (D) at (3, 1);
    \coordinate (E) at (1.75, 1.75);
    \coordinate (F) at (0.5, 2);
    \coordinate (G) at (-1, 1.25);
    \coordinate (H) at (-0.5, 0.25);

    \draw[thick] (A) -- (B) -- (C) -- (D) -- (E) -- (F) -- (G) -- (H) -- (A);

    \node[circle, fill, inner sep = 1.5pt] at (A) {};
    \node[circle, fill, inner sep = 1.5pt] at (B) {};
    \node[circle, fill, inner sep = 1.5pt] at (C) {};
    \node[circle, fill, inner sep = 1.5pt] at (D) {};
    \node[circle, fill, inner sep = 1.5pt] at (E) {};
    \node[circle, fill, inner sep = 1.5pt] at (F) {};
    \node[circle, fill, inner sep = 1.5pt] at (G) {};
    \node[circle, fill, inner sep = 1.5pt] at (H) {};

    \draw[thick, orange, fill=orange!20] (A) -- (C) -- (F) -- (A);

    \coordinate (X) at (0.7, 0.8);
    \coordinate (Y) at (2, 1.1);
    \coordinate (Z) at ($(X)!0.66!(Y)$);

    \draw[thick, blue] (X) -- (Y);
    \node[circle, fill=darkgreen, inner sep=1.5pt, label={[label distance=0mm]left:$\color{darkgreen} x$}] at (X) {};
    \node[circle, fill=darkgreen, inner sep=1.5pt, label={[label distance=0mm]right:$\color{darkgreen} y$}] at (Y) {};
    \node[circle, fill=orange, inner sep=1.5pt, label={[label distance=0mm]above:$\color{orange} z$}] at (Z) {};

    \node[circle, fill, inner sep=1.5pt, label={[label distance=0mm]below:$a$}] at (A) {};
    \node[circle, fill, inner sep=1.5pt, label={[label distance=0mm]below:$b$}] at (B) {};
    \node[circle, fill, inner sep=1.5pt, label={[label distance=0mm]below right:$c$}] at (C) {};
    \node[circle, fill, inner sep=1.5pt, label={[label distance=0mm]right:$d$}] at (D) {};
    \node[circle, fill, inner sep=1.5pt, label={[label distance=0mm]above:$e$}] at (E) {};
    \node[circle, fill, inner sep=1.5pt, label={[label distance=0mm]above:$f$}] at (F) {};
    \node[circle, fill, inner sep=1.5pt, label={[label distance=0mm]left:$g$}] at (G) {};
    \node[circle, fill, inner sep=1.5pt, label={[label distance=0mm]below:$h$}] at (H) {};
\end{tikzpicture}
}

\newcommand{\goodlinepathconvextriangle}[1]{
\begin{tikzpicture}[scale=#1]
    \coordinate (A) at (0, 0);
    \coordinate (B) at (2, 0);
    \coordinate (C) at (3, 0);
    \coordinate (D) at (1, 2);

    \draw[thick] (A) -- (C) -- (D) -- (A);
    \draw[thick, orange] (B) -- (D);
    \node[circle, fill=black, inner sep=1.5pt, label={[label distance=0mm]below left:$a$}] at (A) {};
    \node[circle, fill=black, inner sep=1.5pt, label={[label distance=0mm]below:$b$}] at (B) {};
    \node[circle, fill=black, inner sep=1.5pt, label={[label distance=0mm]below right:$c$}] at (C) {};
    \node[circle, fill=black, inner sep=1.5pt, label={[label distance=0mm]above:$d$}] at (D) {};
\end{tikzpicture}
}

\newcommand{\pocket}[1]{
\begin{tikzpicture}[scale=#1]
    \coordinate (X1) at (0, 0);
    \coordinate (X2) at (0.5, 0.75);
    \coordinate (X3) at (0, 1);
    \coordinate (X4) at (-0.7, 0.7);
    \coordinate (X5) at (0.25, 1.75);
    \coordinate (X6) at (1.3, 0.9);

    \coordinate (X7) at (1.5, 2.1);
    \coordinate (X8) at (0.9, 1.8);
    \coordinate (X9) at (0, 2.75);
    \coordinate (X10) at (-1.5, 2.5);
    \coordinate (X11) at (-0.75, 2);
    \coordinate (X12) at (-0.9, 1.2);
    \coordinate (X13) at (-1.75, 0.25);

    \fill[blue!20] (X1) -- (X2) -- (X3) -- (X4) -- (X5)
        -- (X6) -- (X7) -- (X8) -- (X9) -- (X10) -- (X11) -- (X12) -- (X13) -- cycle;

    \fill[orange!20] (X1) -- (X2) -- (X3) -- (X4) -- (X5) -- (X6) -- cycle;

    \draw[thick, darkgreen] (X1) -- (X2) -- (X3) -- (X4) -- (X5) -- (X6);
    \draw[thick, blue] (X6) -- (X7) -- (X8) -- (X9) -- (X10) -- (X11) -- (X12) -- (X13) -- (X1);
    \draw[thick, orange] (X1) -- (X6);
    \draw[thick, dotted, orange] (X7) -- (X9);
    \draw[thick, dotted, orange] (X10) -- (X13);

    \foreach \i in {1,...,13} {
        \node[circle, fill=black, inner sep=1.5pt] at (X\i) {};
    }

    \node[label={[label distance=-2mm] below right: $v_1$}] at (X1) {};
    \node[label={[label distance=-1mm] above: $v_2$}] at (X2) {};
    \node[label={[label distance=-1mm] below: $v_3$}] at (X3) {};
    \node[label={[label distance=-1mm] below: $v_4$}] at (X4) {};
    \node[label={[label distance=-2mm] above left: $v_5$}] at (X5) {};
    \node[label={[label distance=-2mm] below right: $v_6$}] at (X6) {};
    \node[label={[label distance=-1mm] right: $v_7$}] at (X7) {};
    \node[label={[label distance=-1mm] below: $v_8$}] at (X8) {};
    \node[label={[label distance=-1mm] above: $v_9$}] at (X9) {};
    \node[label={[label distance=-1mm] above: $v_{10}$}] at (X10) {};
    \node[label={[label distance=-1mm, anchor=230]: $v_{11}$}] at (X11) {};
    \node[label={[label distance=-2mm, anchor=210]: $v_{12}$}] at (X12) {};
    \node[label={[label distance=-1mm] below: $v_{13}$}] at (X13) {};
\end{tikzpicture}
}

\newcommand{\badfillinglinepathA}[1]{
\begin{tikzpicture}[scale=#1]
    \coordinate (X1) at (0, 0);
    \coordinate (X2) at (0.25, 0.25);
    \coordinate (X3) at (0.75, 0.25);
    \coordinate (X4) at (1, 0);

    \coordinate (Y1) at (0.75, -0.25);
    \coordinate (Y2) at (0.5, 0);
    \coordinate (Y3) at (0.75, -0.5);
    \coordinate (Y4) at (1.7, 0.6);
    \coordinate (Y5) at (-0.25, 0.6);

    \draw[thick, darkgreen] (X1) -- (X2) -- (X3) -- (X4);
    \draw[thick, orange] (X1) -- (X4);
    \draw[thick, blue] (X4) -- (Y1) -- (Y2) -- (Y3) -- (Y4) -- (Y5) -- (X1);

    \foreach \i in {1,...,4} {
        \node[circle, fill=black, inner sep=1.5pt] at (X\i) {};
    }
    \foreach \i in {1,...,5} {
        \node[circle, fill=black, inner sep=1.5pt] at (Y\i) {};
    }

    \node[label={[label distance=-2mm] below left: $a$}] at (X1) {};
    \node[label={[label distance=-2mm] above right: $b$}] at (X4) {};

    \node[label={[label distance=-2mm] above left: $x_1$}] at (X2) {};
    \node[label={[label distance=-2mm] above right: $x_2$}] at (X3) {};
\end{tikzpicture}
}

\newcommand{\badfillinglinepathB}[1]{
\begin{tikzpicture}[scale=#1]
    \coordinate (X1) at (0, 0);
    \coordinate (X2) at (0.25, 0.5);
    \coordinate (X3) at (0.75, 0.5);
    \coordinate (X4) at (1, 0);

    \coordinate (Y1) at (0.75, 0.25);
    \coordinate (Y2) at (0.5, 0);
    \coordinate (Y3) at (0.25, 0.25);

    \draw[thick, darkgreen] (X1) -- (X2) -- (X3) -- (X4);
    \draw[thick, orange] (X1) -- (X4);
    \draw[thick, blue] (X4) -- (Y1) -- (Y2) -- (Y3) -- (X1);

    \foreach \i in {1,...,4} {
        \node[circle, fill=black, inner sep=1.5pt] at (X\i) {};
    }
    \foreach \i in {1,...,3} {
        \node[circle, fill=black, inner sep=1.5pt] at (Y\i) {};
    }

    \node[label={[label distance=0mm] left: $a$}] at (X1) {};
    \node[label={[label distance=0mm] right: $b$}] at (X4) {};

    \node[label={[label distance=-2mm] above left: $x_1$}] at (X2) {};
    \node[label={[label distance=-2mm] above right: $x_2$}] at (X3) {};
\end{tikzpicture}
}

\newcommand{\badfillinglinepathC}[1]{
\begin{tikzpicture}[scale=#1]
    \coordinate (X1) at (0, 0);
    \coordinate (X2) at (0.25, 0.5);
    \coordinate (X3) at (0.75, 0.5);
    \coordinate (X4) at (1, 0);

    \coordinate (Y1) at (0.7, 0.25);
    \coordinate (Y2) at (0.4, 0);
    \coordinate (Y3) at (0.8, 0.8);
    \coordinate (Y4) at (-0.2, 0.8);

    \draw[thick, darkgreen] (X1) -- (X2) -- (X3) -- (X4);
    \draw[thick, orange] (X1) -- (X4);
    \draw[thick, blue] (X4) -- (Y1) -- (Y2) -- (Y3) -- (Y4) -- (X1);

    \foreach \i in {1,...,4} {
        \node[circle, fill=black, inner sep=1.5pt] at (X\i) {};
    }
    \foreach \i in {1,...,4} {
        \node[circle, fill=black, inner sep=1.5pt] at (Y\i) {};
    }

    \node[label={[label distance=0mm] left: $a$}] at (X1) {};
    \node[label={[label distance=0mm] right: $b$}] at (X4) {};

    \node[label={[label distance=-2mm] above left: $x_1$}] at (X2) {};
    \node[label={[label distance=-1mm] right: $x_2$}] at (X3) {};
\end{tikzpicture}
}

\newcommand{\badfillinglinepathD}[1]{
\begin{tikzpicture}[scale=#1]
    \coordinate (X1) at (0, 0);
    \coordinate (X2) at (0.25, 0.25);
    \coordinate (X3) at (0.5, 0);
    \coordinate (X4) at (0.75, -0.25);
    \coordinate (X5) at (1, 0);

    \coordinate (Y1) at (0.75, -0.7);
    \coordinate (Y2) at (1.5, 0.2);
    \coordinate (Y3) at (0.25, 0.5);

    \draw[thick, darkgreen] (X1) -- (X2) -- (X3) -- (X4) -- (X5);
    \draw[thick, orange] (X1) -- (X5);
    \draw[thick, blue] (X5) -- (Y1) -- (Y2) -- (Y3) -- (X1);

    \foreach \i in {1,...,5} {
        \node[circle, fill=black, inner sep=1.5pt] at (X\i) {};
    }
    \foreach \i in {1,...,3} {
        \node[circle, fill=black, inner sep=1.5pt] at (Y\i) {};
    }

    \node[label={[label distance=0mm] left: $a$}] at (X1) {};
    \node[label={[label distance=-2mm] above right: $b$}] at (X5) {};

    \node[label={[label distance=-1mm] right: $x_1$}] at (X2) {};
    \node[label={[label distance=-2.5mm] below left: $x_3$}] at (X4) {};
\end{tikzpicture}
}

\newcommand{\badfillinglinepathE}[1]{
\begin{tikzpicture}[scale=#1]
    \coordinate (X1) at (0, 0);
    \coordinate (X2) at (0.25, 0.25);
    \coordinate (X3) at (0.5, 0);
    \coordinate (X4) at (0.75, 0.25);
    \coordinate (X5) at (1, 0);

    \coordinate (Y1) at (0.75, 0.5);
    \coordinate (Y2) at (0.25, 0.5);

    \draw[thick, darkgreen] (X1) -- (X2) -- (X3) -- (X4) -- (X5);
    \draw[thick, orange] (X1) -- (X5);
    \draw[thick, blue] (X5) -- (Y1) -- (Y2) -- (X1);

    \foreach \i in {1,...,5} {
        \node[circle, fill=black, inner sep=1.5pt] at (X\i) {};
    }
    \foreach \i in {1,...,2} {
        \node[circle, fill=black, inner sep=1.5pt] at (Y\i) {};
    }

    \node[label={[label distance=0mm] left: $a$}] at (X1) {};
    \node[label={[label distance=0mm] right: $b$}] at (X5) {};

    \node[label={[label distance=-1mm] below: $x_2$}] at (X3) {};
\end{tikzpicture}
}

\newcommand{\badfillinglinepathF}[1]{
\begin{tikzpicture}[scale=#1]
    \coordinate (X1) at (0, 0);
    \coordinate (X2) at (0.25, 0.25);
    \coordinate (X3) at (0.5, 0);
    \coordinate (X4) at (0.75, -0.25);
    \coordinate (X5) at (1, 0);

    \coordinate (Y1) at (1.23, -0.5);
    \coordinate (Y2) at (0.75, -0.5);
    \coordinate (Y3) at (0.5, -0.25);
    \coordinate (Y4) at (1, 0.25);
    \coordinate (Y5) at (0.25, 0.5);

    \draw[thick, darkgreen] (X1) -- (X2) -- (X3) -- (X4) -- (X5);
    \draw[thick, orange] (X1) -- (X5);
    \draw[thick, blue] (X5) -- (Y1) -- (Y2) -- (Y3) -- (Y4) -- (Y5) -- (X1);

    \foreach \i in {1,...,5} {
        \node[circle, fill=black, inner sep=1.5pt] at (X\i) {};
    }
    \foreach \i in {1,...,5} {
        \node[circle, fill=black, inner sep=1.5pt] at (Y\i) {};
    }

    \node[label={[label distance=0mm] left: $a$}] at (X1) {};
    \node[label={[label distance=0mm] right: $b$}] at (X5) {};

    \node[label={[label distance=-1mm] right: $x_1$}] at (X2) {};
    \node[label={[label distance=-1mm] right: $x_3$}] at (X4) {};
\end{tikzpicture}
}

\newcommand{\subpathintersectionA}[1]{
\begin{tikzpicture}[scale=#1]
    \coordinate (X1) at (0, 0);
    \coordinate (X2) at (-0.5, 0.25);
    \coordinate (X3) at (0, 0.5);
    \coordinate (X4) at (-0.25, 0.75);
    \coordinate (X5) at (0.25, 0.75);
    \coordinate (X6) at (0.75, 0.3);
    \coordinate (X7) at (2, 0.75);
    \coordinate (X8) at (2.5, 0.5);
    \coordinate (X9) at (1.5, 0.1);
    \coordinate (X10) at (1.8, -0.4);
    \coordinate (X11) at (2.3, -0.2);
    \coordinate (X12) at (2, 0);

    \coordinate (Yr) at (0.75, -0.5);
    \coordinate (Ys) at (1.1, 0.9);

    \coordinate (Y1) at (1.2, -0.25);
    \coordinate (Y2) at (-0.25, 0.25);
    \coordinate (Y3) at (0.25, 0.5);
    \coordinate (Y4) at (0.75, 0.15);
    \coordinate (Y5) at (1.8, 0.5);

    \coordinate (R1) at (3, 0);
    \coordinate (R2) at (3, 1.4);
    \coordinate (R3) at (-1, 1.4);
    \coordinate (R4) at (-1, 0);

    \coordinate (X) at (1.1, 1.4);
    \coordinate (Z) at (1.2, 1.25);

    \draw[thick, orange] (X1) -- (X12);
    \draw[thick, darkgreen] (X1) -- (X2) -- (X3) -- (X4) -- (X5) -- (X6) -- (X7) -- (X8) -- (X9) -- (X10) -- (X11) -- (X12);
    \draw[thick, dashed] (X12) -- (R1) -- (R2) -- (R3) -- (R4) -- (X1);

    \node[label={[label distance=0mm] below: $a$}] at (X1) {};
    \node[label={[label distance=-1mm] below left: $b$}] at (X12) {};

    \node[circle, fill=black, inner sep=1pt, label={[label distance=0mm] below right: $r_1$}] at (R1) {};
    \node[circle, fill=black, inner sep=1pt, label={[label distance=0mm] above right: $r_2$}] at (R2) {};
    \node[circle, fill=black, inner sep=1pt, label={[label distance=0mm] above left: $r_3$}] at (R3) {};
    \node[circle, fill=black, inner sep=1pt, label={[label distance=0mm] below left: $r_4$}] at (R4) {};

    \draw[orange] (Ys) -- (Z);

    \draw[thick, blue, dotted] (Yr) -- (Y1) -- (Y2) -- (Y3) -- (Y4) -- (Y5) -- (Ys);

    \draw[red, dashed] (X) circle (0.3) {};

    \draw (Yr) edge[red, dashed, ->] (0.75, -1);

    \node[circle, fill=black, inner sep=1pt] at (X) {};

    \node[circle, fill=orange, inner sep=1pt, label={[label distance=-0.5mm] left: $\color{orange} z$}] at (Z) {};

    \node[circle, fill=black, inner sep=1.5pt, label={[label distance=-1mm, anchor=150]: $y_r$}] at (Yr) {};
    \node[circle, fill=black, inner sep=1.5pt, label={[label distance=-2mm] below left: $y_s$}] at (Ys) {};

    \node[circle, fill=black, inner sep=1.5pt] at (X1) {};
    \node[circle, fill=black, inner sep=1.5pt] at (X12) {};
    \foreach \i in {2,...,11} {
        \node[circle, fill=black, inner sep=1pt] at (X\i) {};
    }

    \foreach \i in {1,...,5} {
        \node[circle, fill=blue, inner sep=1pt] at (Y\i) {};
    }
\end{tikzpicture}
}

\newcommand{\subpathintersectionB}[1]{
\begin{tikzpicture}[scale=#1]
    \coordinate (X1) at (0, 0);
    \coordinate (X2) at (-0.5, 0.25);
    \coordinate (X3) at (0, 0.5);
    \coordinate (X4) at (-0.25, 0.75);
    \coordinate (X5) at (0.25, 0.75);
    \coordinate (X6) at (0.75, 0.3);
    \coordinate (X7) at (2, 0.75);
    \coordinate (X8) at (2.5, 0.5);
    \coordinate (X9) at (1.7, 0.3);
    \coordinate (X10) at (2, 0);

    \coordinate (Yr) at (0.5, 0);
    \coordinate (Ys) at (1.5, 1);

    \coordinate (Y1) at (-0.25, 0.25);
    \coordinate (Y2) at (0.25, 0.5);
    \coordinate (Y3) at (0.75, 0.15);
    \coordinate (Y4) at (1.8, 0.5);

    \coordinate (R1) at (2, -0.5);
    \coordinate (R2) at (0, -0.5);

    \coordinate (X) at (1.5, 1.5);
    \coordinate (Z) at (1.6, 1.3);

    \draw[thick, orange] (X1) -- (X10);
    \draw[thick, darkgreen] (X1) -- (X2) -- (X3) -- (X4) -- (X5) -- (X6) -- (X7) -- (X8) -- (X9) -- (X10);
    \draw[thick, dashed] (X10) -- (R1) -- (R2) -- (X1);

    \node[label={[label distance=-1mm] below left: $a$}] at (X1) {};
    \node[label={[label distance=0mm] right: $b$}] at (X10) {};

    \node[circle, fill=black, inner sep=1pt, label={[label distance=0mm] below right: $r_1$}] at (R1) {};
    \node[circle, fill=black, inner sep=1pt, label={[label distance=0mm] below left: $r_2$}] at (R2) {};

    \draw[thick, blue, dotted] (Yr) -- (Y1) -- (Y2) -- (Y3) -- (Y4) -- (Ys);

    \node[circle, fill=black, inner sep=1.5pt, label={[label distance=0mm] below: $y_r$}] at (Yr) {};
    \node[circle, fill=black, inner sep=1.5pt, label={[label distance=0mm] left: $y_s$}] at (Ys) {};

    \node[circle, fill=black, inner sep=1.5pt] at (X1) {};
    \node[circle, fill=black, inner sep=1.5pt] at (X10) {};
    \foreach \i in {2,...,9} {
        \node[circle, fill=black, inner sep=1pt] at (X\i) {};
    }

    \foreach \i in {1,...,4} {
        \node[circle, fill=blue, inner sep=1pt] at (Y\i) {};
    }
\end{tikzpicture}
}

\newcommand{\pocketpathgood}[1]{
\begin{tikzpicture}[scale=#1]
    \coordinate (X1) at (0, 0);
    \coordinate (X2) at (-0.5, 0.25);
    \coordinate (X3) at (0, 0.5);
    \coordinate (X4) at (-0.25, 0.75);
    \coordinate (X5) at (0.25, 0.9);
    \coordinate (X6) at (0.9, 0.7);
    \coordinate (X7) at (2, 1.25);
    \coordinate (X8) at (2.9, 0.4);
    \coordinate (X9) at (1.6, 0.8);
    \coordinate (X10) at (2, 0);

    \coordinate (Y2) at (2.5, 0);
    \coordinate (Y3) at (4, 0.7);
    \coordinate (Y4) at (3.8, 1.5);
    \coordinate (Y5) at (2.7, 2);
    \coordinate (Y6) at (3.5, 0.7);
    \coordinate (Y7) at (1.5, 1.7);
    \coordinate (Y8) at (1, 1);
    \coordinate (Y9) at (0.35, 1.2);
    \coordinate (Y10) at (-0.7, 0.9);
    \coordinate (Y11) at (-0.35, 0.52);
    \coordinate (Y12) at (-1.1, 0.2);

    \coordinate (center) at (1.2, 0);
    \coordinate (X') at (3.8, 1.9);
    \coordinate (Z) at (1.4, 0.3);
    \coordinate (Z') at (1, 0.2);
    \coordinate (X) at ($(Z)!0.52!(X')$);
    \coordinate(y) at ($(Z)!0.24!(X')$);
    \coordinate(e) at ($(Z)!0.57!(X)$);

    \draw[thick, orange] (X1) -- (X10);
    \draw[thick, darkgreen] (X1) -- (X2) -- (X3) -- (X4) -- (X5) -- (X6) -- (X7) -- (X8) -- (X9) -- (X10);
    \draw[thick, blue] (X10) -- (Y2) -- (Y3) -- (Y4) -- (Y5) -- (Y6) -- (Y7) -- (Y8) -- (Y9) -- (Y10) -- (Y11) -- (Y12) -- (X1);

    \draw[red, dotted, thick] (Z) -- (Z') {};

    \node[label={[label distance=-1mm] below: $a$}] at (X1) {};
    \node[label={[label distance=0mm] below: $b$}] at (X10) {};

    \draw[thick, orange] (Z) edge[->] (X');

    \draw[red, dashed] (center) circle (0.6) {};

    \node[circle, fill=black, inner sep=0.8pt] at (center) {};
    \node[circle, fill=orange, inner sep=0.8pt] at (X) {};
    \node[circle, fill=orange, inner sep=0.8pt, label={[label distance=-1mm] below right: $\color{orange} z$}] at (Z) {};
    \node[circle, fill=orange, inner sep=0.8pt, label={[label distance=-1mm] left: $\color{orange} z'$}] at (Z') {};
    \node[circle, fill=purple, inner sep=0.8pt, label={[label distance=-0.5mm, anchor=100]: $\color{purple} y$}] at (y) {};
    \node[circle, fill=purple, inner sep=0.8pt, label={[label distance=0mm] above: $\color{purple} x$}] at (X) {};

    \node[circle, fill=black, inner sep=1.5pt] at (X1) {};
    \node[circle, fill=black, inner sep=1.5pt] at (X10) {};
    \foreach \i in {2,...,9} {
        \node[circle, fill=black, inner sep=1pt] at (X\i) {};
    }

    \foreach \i in {2,...,12} {
        \node[circle, fill=black, inner sep=1pt] at (Y\i) {};
    }
\end{tikzpicture}
}

\newcommand{\polygonalpathparametrization}[1]{
\begin{tikzpicture}[scale=#1]
    \foreach \i in {0,...,5} {
        \coordinate (X\i) at (\i, 0);
        \node[circle, fill=black, inner sep=1pt, label={below: $v_\i$}] at (X\i) {};
    }
    \draw[thick] (X0) -- (X1) -- (X2) -- (X3) -- (X4) -- (X5);
    \draw [decorate,decoration={brace,amplitude=5pt},xshift=0pt,yshift=0pt]
        (X0) -- (X1) node [black,midway,yshift=13pt] {$\frac{1}{2}$};
    \draw [decorate,decoration={brace,amplitude=5pt},xshift=0pt,yshift=0pt]
        (X1) -- (X2) node [black,midway,yshift=13pt] {$\frac{1}{4}$};
    \draw [decorate,decoration={brace,amplitude=5pt},xshift=0pt,yshift=0pt]
        (X2) -- (X3) node [black,midway,yshift=13pt] {$\frac{1}{8}$};
    \draw [decorate,decoration={brace,amplitude=5pt},xshift=0pt,yshift=0pt]
        (X3) -- (X4) node [black,midway,yshift=13pt] {$\frac{1}{16}$};
    \draw [decorate,decoration={brace,amplitude=5pt},xshift=0pt,yshift=0pt]
        (X4) -- (X5) node [black,midway,yshift=13pt] {$\frac{1}{16}$};
\end{tikzpicture}
}

\newcommand{\rref}[2][]{\prettyref{#2}}
\newrefformat{sec}{Sect.\,\ref{#1}}%
\newrefformat{def}{Definition\,\ref{#1}}
\newrefformat{thm}{Theorem\,\ref{#1}}
\newrefformat{prop}{Proposition\,\ref{#1}}
\newrefformat{rem}{Remark\,\ref{#1}}
\newrefformat{lem}{Lemma\,\ref{#1}}
\newrefformat{cor}{Corollary\,\ref{#1}}
\newrefformat{ex}{Example\,\ref{#1}}
\newrefformat{tab}{Table\,\ref{#1}}
\newrefformat{fig}{Fig.\,\ref{#1}}
\newrefformat{case}{case\,\ref{#1}}
\newrefformat{foot}{Footnote\,\ref{#1}}
\newrefformat{app}{Appendix\,\ref{#1}}
\newrefformat{chap}{Chapter\,\ref{#1}}

\newcommand{\tit}{\textit}

\newcommand{\from}{\colon}

\newcommand{\R}{\mathbb{R}}
\newcommand{\Z}{\mathbb{Z}}
\renewcommand{\emptyset}{\varnothing}
\newcommand{\mpp}{\isa{make\_polygonal\_path}}

\newcommand{\snip}[4]{\expandafter\newcommand\csname #1\endcsname{#4}}
\snip{pick}{1}{2}{%
\isacommand{theorem}\isamarkupfalse%
\ pick{\isacharcolon}{\kern0pt}\isanewline
\ \ \isakeyword{assumes}\ {\isachardoublequoteopen}polygon\ p{\isachardoublequoteclose}\ {\isachardoublequoteopen}p\ {\isacharequal}{\kern0pt}\ make{\isacharunderscore}{\kern0pt}polygonal{\isacharunderscore}{\kern0pt}path\ vts{\isachardoublequoteclose}\ {\isachardoublequoteopen}all{\isacharunderscore}{\kern0pt}integral\ vts{\isachardoublequoteclose}\isanewline
\ \ \isakeyword{assumes}\ {\isachardoublequoteopen}I\ {\isacharequal}{\kern0pt}\ card\ {\isacharbraceleft}{\kern0pt}x{\isachardot}{\kern0pt}\ integral{\isacharunderscore}{\kern0pt}vec\ x\ {\isasymand}\ x\ {\isasymin}\ path{\isacharunderscore}{\kern0pt}inside\ p{\isacharbraceright}{\kern0pt}{\isachardoublequoteclose}\ \isanewline
\ \ \isakeyword{assumes}\ {\isachardoublequoteopen}B\ {\isacharequal}{\kern0pt}\ card\ {\isacharbraceleft}{\kern0pt}x{\isachardot}{\kern0pt}\ integral{\isacharunderscore}{\kern0pt}vec\ x\ {\isasymand}\ x\ {\isasymin}\ path{\isacharunderscore}{\kern0pt}image\ p{\isacharbraceright}{\kern0pt}{\isachardoublequoteclose}\isanewline
\ \ \isakeyword{shows}\ {\isachardoublequoteopen}measure\ lebesgue\ {\isacharparenleft}{\kern0pt}path{\isacharunderscore}{\kern0pt}inside\ p{\isacharparenright}{\kern0pt}\ {\isacharequal}{\kern0pt}\ I\ {\isacharplus}{\kern0pt}\ B{\isacharslash}{\kern0pt}{\isadigit{2}}\ {\isacharminus}{\kern0pt}\ {\isadigit{1}}{\isachardoublequoteclose}%
}%
\snip{rotationlemma}{1}{2}{%
\isacommand{lemma}\isamarkupfalse%
\ rotation{\isacharunderscore}{\kern0pt}is{\isacharunderscore}{\kern0pt}polygon{\isacharcolon}{\kern0pt}\isanewline
\ \ \isakeyword{assumes}\ {\isachardoublequoteopen}polygon\ p{\isachardoublequoteclose}\ {\isachardoublequoteopen}p\ {\isacharequal}{\kern0pt}\ make{\isacharunderscore}{\kern0pt}polygonal{\isacharunderscore}{\kern0pt}path\ vts{\isachardoublequoteclose}\isanewline
\ \ \isakeyword{assumes}\ {\isachardoublequoteopen}vts{\isacharprime}{\kern0pt}\ {\isacharequal}{\kern0pt}\ rotate{\isacharunderscore}{\kern0pt}polygon{\isacharunderscore}{\kern0pt}vertices\ vts\ i{\isachardoublequoteclose}\isanewline
\ \ \isakeyword{shows}\ {\isachardoublequoteopen}polygon\ {\isacharparenleft}{\kern0pt}make{\isacharunderscore}{\kern0pt}polygonal{\isacharunderscore}{\kern0pt}path\ vts{\isacharprime}{\kern0pt}{\isacharparenright}{\kern0pt}{\isachardoublequoteclose}\ \isakeyword{and}\isanewline
\ \ \ \ \ \ \ \ {\isachardoublequoteopen}path{\isacharunderscore}{\kern0pt}image\ {\isacharparenleft}{\kern0pt}make{\isacharunderscore}{\kern0pt}polygonal{\isacharunderscore}{\kern0pt}path\ vts{\isacharprime}{\kern0pt}{\isacharparenright}{\kern0pt}\ {\isacharequal}{\kern0pt}\ path{\isacharunderscore}{\kern0pt}image\ p{\isachardoublequoteclose}%
}%
\snip{polygondef}{1}{2}{%
\isacommand{definition}\isamarkupfalse%
\ polygon\ {\isacharcolon}{\kern0pt}{\isacharcolon}{\kern0pt}\ {\isachardoublequoteopen}{\isacharparenleft}{\kern0pt}real\ {\isasymRightarrow}\ real{\isacharcircum}{\kern0pt}{\isadigit{2}}{\isacharparenright}{\kern0pt}\ {\isasymRightarrow}\ bool{\isachardoublequoteclose}\ \isakeyword{where}\isanewline
\ \ {\isachardoublequoteopen}polygon\ g\ {\isasymlongleftrightarrow}\ polygonal{\isacharunderscore}{\kern0pt}path\ g\ {\isasymand}\ simple{\isacharunderscore}{\kern0pt}path\ g\ {\isasymand}\ closed{\isacharunderscore}{\kern0pt}path\ g{\isachardoublequoteclose}%
}%
\snip{mppdef}{1}{2}{%
\isacommand{fun}\isamarkupfalse%
\ make{\isacharunderscore}{\kern0pt}polygonal{\isacharunderscore}{\kern0pt}path\ {\isacharcolon}{\kern0pt}{\isacharcolon}{\kern0pt}\ {\isachardoublequoteopen}{\isacharparenleft}{\kern0pt}real{\isacharcircum}{\kern0pt}{\isadigit{2}}{\isacharparenright}{\kern0pt}\ list\ {\isasymRightarrow}\ {\isacharparenleft}{\kern0pt}real\ {\isasymRightarrow}\ real{\isacharcircum}{\kern0pt}{\isadigit{2}}{\isacharparenright}{\kern0pt}{\isachardoublequoteclose}\ \isakeyword{where}\isanewline
\ \ {\isachardoublequoteopen}make{\isacharunderscore}{\kern0pt}polygonal{\isacharunderscore}{\kern0pt}path\ {\isacharbrackleft}{\kern0pt}{\isacharbrackright}{\kern0pt}\ {\isacharequal}{\kern0pt}\ linepath\ {\isadigit{0}}\ {\isadigit{0}}{\isachardoublequoteclose}\isanewline
{\isacharbar}{\kern0pt}\ {\isachardoublequoteopen}make{\isacharunderscore}{\kern0pt}polygonal{\isacharunderscore}{\kern0pt}path\ {\isacharbrackleft}{\kern0pt}a{\isacharbrackright}{\kern0pt}\ {\isacharequal}{\kern0pt}\ linepath\ a\ a{\isachardoublequoteclose}\isanewline
{\isacharbar}{\kern0pt}\ {\isachardoublequoteopen}make{\isacharunderscore}{\kern0pt}polygonal{\isacharunderscore}{\kern0pt}path\ {\isacharbrackleft}{\kern0pt}a{\isacharcomma}{\kern0pt}b{\isacharbrackright}{\kern0pt}\ {\isacharequal}{\kern0pt}\ linepath\ a\ b{\isachardoublequoteclose}\isanewline
{\isacharbar}{\kern0pt}\ {\isachardoublequoteopen}make{\isacharunderscore}{\kern0pt}polygonal{\isacharunderscore}{\kern0pt}path\ {\isacharparenleft}{\kern0pt}a\ {\isacharhash}{\kern0pt}\ b\ {\isacharhash}{\kern0pt}\ xs{\isacharparenright}{\kern0pt}\ {\isacharequal}{\kern0pt}\isanewline
\ \ \ \ {\isacharparenleft}{\kern0pt}linepath\ a\ b{\isacharparenright}{\kern0pt}\ {\isacharplus}{\kern0pt}{\isacharplus}{\kern0pt}{\isacharplus}{\kern0pt}\ make{\isacharunderscore}{\kern0pt}polygonal{\isacharunderscore}{\kern0pt}path\ {\isacharparenleft}{\kern0pt}b\ {\isacharhash}{\kern0pt}\ xs{\isacharparenright}{\kern0pt}{\isachardoublequoteclose}%
}%

\makeatletter
\def\orcidID#1{\href{http://orcid.org/#1}{\protect\raisebox{-1.25pt}{\protect\includegraphics{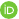}}}}
\makeatother
\begin{document}
\title{Formalizing Pick's Theorem in Isabelle/HOL}
\author{Sage Binder\inst{1}\orcidID{0009-0004-4776-2018} \and
Katherine Kosaian\inst{2}\orcidID{0000-0002-9336-6006}
\authorrunning{S. Binder and K. Kosaian}
\institute{University of Iowa, Iowa City IA 52242, USA\\
\email{sage-binder@uiowa.edu}\\
\and
Iowa State University, Ames IA 50011, USA\\
\email{kkosaian@iastate.edu}}
}
\maketitle              %
\begin{abstract}
We formalize Pick’s theorem for finding the area of a simple polygon whose vertices are integral lattice points. We are inspired by John Harrison's formalization of Pick's theorem in HOL Light, but tailor our proof approach to avoid a primary challenge point in his formalization, which is proving that any polygon with more than three vertices can be split (in its interior) by a line between some two vertices. We detail the approach we use to avoid this step and reflect on the pros and cons of our eventual formalization strategy.
We use the theorem prover Isabelle/HOL, and our formalization involves augmenting the existing geometry libraries in various foundational ways (e.g., by adding the definition of a polygon and formalizing some key properties thereof). 

\keywords{Pick's theorem \and Isabelle/HOL \and formalization \and geometry.}
\end{abstract}
\section{Introduction}

Pick's theorem is a gem of late nineteenth-century mathematics, first proved by George Pick in 1899.
It concerns simple polygons (i.e., polygons with no self-intersections or holes) whose vertices are integral lattice points and gives a simple-to-state relationship between a polygon's area and the number of integral lattice points within and on the boundary of the polygon.
Pick's theorem has a number of interesting applications
in geometry as well as in number theory \cite{pickapplications},
but the main reason for its beloved status is arguably the elegant simplicity
of the theorem statement.
Perhaps for this reason,
it was included on the list of ``Top 100 Theorems'' compiled by Paul and Jack Abad,
and is currently tracked in Freek Wiedijk's corresponding list of
``Formalizing 100 Theorems'' \cite{freek},
which tracks which of these theorems have been formalized in (some of the major) theorem provers.
In this list,
only one theorem---Fermat's Last Theorem---has not been formalized in any theorem prover,
and most have been formalized in multiple provers.
Pick's theorem is one of the few listed theorems which have only been
formalized in one prover,
namely HOL Light (by John Harrison) \cite{DBLP:journals/mscs/Harrison11}.

We suspect that the inherent challenge in formalizing geometry has something to do with why
Pick's theorem has only been formalized in one theorem prover.
Harrison comments on this challenge in his work \cite{DBLP:journals/mscs/Harrison11}.
Informal geometric arguments often rely on proof-by-picture intuition,
and formalizing these arguments involves working with abstract definitions
which are often unwieldy and rather removed from geometric intuition.
Consequently,
it is not uncommon for the geometry libraries of a theorem prover to be less well-developed
than libraries for other areas of mathematics, such as algebra and analysis.

We formalize Pick's theorem in Isabelle/HOL \cite{DBLP:journals/jar/Paulson89,DBLP:books/sp/NipkowPW02}.
Our work is inspired by Harrison's,
but motivated by an interest
in avoiding the challenge point of proving
that any polygon with more than three vertices can be split into two polygons
by a line between two of its vertices.
Harrison identifies this step as being particularly arduous,
and comments that the proof of Pick's theorem is significantly simpler for convex polygons
\cite{DBLP:journals/mscs/Harrison11}.
Our approach (discussed in \rref{sec:proof})
splits the proof into the convex and non-convex cases;
in the non-convex case,
we construct a path between two vertices of the polygon
that lies fully \textit{outside} the polygon.
In the process,
we make various contributions to the Isabelle libraries,
such as the formal definition of a polygon and properties of polygons
(formalization details are discussed in \rref{sec:formalization}). 
Ultimately,
we encounter difficulties similar to those discussed by Harrison,
and thus do not claim that our approach is simpler than his
(we give a retrospective analysis in the conclusion, \rref{sec:conclusion}).
Nevertheless,
our proof is a novel formalization of Pick's theorem,
and our treatment of the non-convex case involves various creative steps.

Isabelle/HOL,
our theorem prover of choice,
is well-suited to formalizing mathematics.
Our work is facilitated by Isabelle's automated proof search,
particularly Sledgehammer \cite{DBLP:conf/lpar/PaulsonB10},
and existing library search tools,
particularly SErAPIS \cite{stathopoulos2020serapis}.
We simultaneously benefit from key results that are already formalized in Isabelle's libraries
and suffer from the library's foundational gaps.
For example,
we use the Jordan curve theorem and a variant,
called the Jordan triple curve theorem,
which are already proved in Isabelle/HOL
(and, further, have been specialized to our setting of $\mathbb{R}^2$).
On the other hand,
we did not find an existing notion of a polygon in the libraries.

Our formalization is about $14300$ lines of code,
and is available on the Archive of Formal Proofs (AFP) \cite{Picks_Theorem-AFP}.

\section{Related Work} \label{sec:related}
In Isabelle/HOL's Archive of Formal Proofs (AFP)\footnote{\url{https://www.isa-afp.org}},
which is a large centralized repository of proof developments,
there are 22 entries categorized under
``Mathematical Geometry''\footnote{\url{https://www.isa-afp.org/topics/mathematics/geometry/}}
(some entries with a geometric flavor  are categorized elsewhere,
e.g. a recent formalization \cite{DBLP:conf/cpp/ImmlerT20} of the Poincar\'{e}-Bendixson theorem
which required formalizing proofs-by-picture is categorized under ``Analysis'').
Notably, Isabelle's libraries already contain some results about triangles---in particular,
there is a library containing some basic properties of triangles \cite{Triangle-AFP},
and further results have been proved on top of this, including the intersecting chords theorem
\cite{Chord_Segments-AFP}, Stewart's theorem \cite{Stewart_Apollonius-AFP},
and most recently Ceva's theorem \cite{Ceva-AFP}.
The underlying triangle library \cite{Triangle-AFP}
defines a triangle as three points in a real inner product space,
but in our setting,
it is more natural to define triangles (and general polygons)
in terms of their boundary.
The formalization of Ceva's theorem \cite{Ceva-AFP}
contains a notion of area for triangles,
but it is defined in terms of the side lengths of the triangle and the measure of one of the angles.
In our formalization,
we instead treat area as the Lebesgue measure of the inside of a polygon.
Outside of the AFP,
implementing (but not verifying) an algorithm to triangulate a polygon was the subject
of a recent master's thesis \cite{skjelnes2020implementation}.

Among theorem provers,
HOL Light has particularly well-developed geometry libraries \cite{DBLP:journals/jar/Harrison13},
and a portion of the geometry results formalized
in Isabelle have been ported from HOL Light,
including the Jordan curve theorem
(formalized by Hales in 2007 \cite{DBLP:journals/tamm/Hales07}
and later ported to Isabelle by Paulson),
and Euler's Polyhedron Formula
(ported to Isabelle in 2023 by Paulson \cite{Euler_Polyhedron_Formula-AFP}).
Some Euclidean geometry is formalized in Lean's MathLib,
including properties of triangles similar to those in Isabelle \cite{Lean_Mathlib_Triangles};
in 2022 Myers formalized the solution to a 2019 geometric IMO problem in Lean \cite{Lean_Mathlib_IMO}.
In Coq, some algorithms for triangulating convex hulls have been formalized \cite{DBLP:conf/ictac/Bertot18,DBLP:conf/itp/DufourdB10}, and
many results in Euclidean geometry founded on Tarski's geometry axioms
have been formalized in the GeoCoq library \cite{geocoq}
(which has been partially ported to Isabelle/HOL \cite{IsaGeoCoq-AFP}).

\section{Our Proof Approach}\label{sec:proof}
\begin{figure}
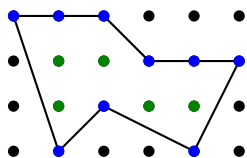
 
\centering
    \pickexample{0.6}
    \caption{We visualize the five lattice points inside the polygon in green and the nine points on the boundary in blue, and the area of the polygon is $5 + 9/2 -1 = 8.5$.} \label{fig:Pick}
\end{figure}

Pick's theorem \cite{pick1899geometrisches} says that
the area of a simple polygon whose vertices are integral lattice points
is equal to the number of integral lattice points inside the polygon,
plus half the number of integral lattice points on the boundary,
minus $1$ (see \rref{fig:Pick}).
The standard proof of Pick's theorem proceeds by (strong) induction
on the number of vertices of a polygon,
and involves splitting the polygon into two smaller polygons in the inductive step.
While we largely follow this standard proof (which is also the approach that Harrison takes),
we sidestep the need to prove that any polygon with more than three vertices
can be split,
so as to avoid a number of painful challenges \cite{DBLP:journals/mscs/Harrison11}
which Harrison encountered.

The high-level structure in our proof is as follows.
Let $p$ be a polygon. %
We proceed by strong induction on the number of vertices
defining $p$. %
In the base case, we prove Pick's theorem for triangles
(polygons with three vertices).
In the inductive case, where the polygon has more than three vertices,
we depart from Harrison's approach by splitting into two subcases:
when $p$ is convex,
and when $p$ is not convex.
We choose to case on convexity since
it is mathematically trivial to show that
a convex polygon (with more than three vertices)
can be split into two smaller polygons on which we can induct.
In the non-convex case,
we use the convex hull of the polygon to find a linepath lying entirely outside the polygon,
which avoids the need to split an arbitrary non-convex polygon.
This avoids specific challenges Harrison faced in his formalization,
but also presents new challenges of a similar flavor, which we solve with a new approach.

We first discuss some preliminaries (\rref{sec:prelim}) and then further detail our approach (\rref{sec:approach-base-case}---\rref{sec:approach-non-convex}), focusing on the novel aspects of our work, particularly the techniques we use in the non-convex case (\rref{sec:approach-non-convex}).
We present our top-level result in \rref{sec:toplevel}.

\subsection{Preliminaries}\label{sec:prelim}
Formally stating Pick's theorem requires a formal definition of a polygon
as well as notions of area, boundary, and inside.
Auspiciously,
many of the core geometric definitions used in Harrison's formalization of Pick's theorem
were already present in Isabelle/HOL's libraries.
\paragraph{Paths and Convex Hulls.}
We build on various properties from the libraries about paths.
The library definition of a path is a continuous mapping from the interval $[0, 1]$ into some topological space;
in our setting,
we care about paths in $\R^2$.
The libraries provide definitions and lemmas for simple paths,%
\footnote{A path which does not intersect itself except potentially at its endpoints.}
linepaths,%
\footnote{A straight line between two points.}
and joining paths together.
In this paper,
we frequently refer to the \tit{path image} of a given path $p$,
which is defined in the libraries as the image of $[0, 1]$ under $p$.

The libraries also provide convex hull results,
where the convex hull of a set $S$ is defined as the minimal convex set containing all of $S$.
Additionally,
the libraries provide the definition of an \tit{extreme point} of a set;
intuitively,
for a convex hull $H$ of a finite set,
an extreme point of $H$ is a ``corner'' of $H$.
One particularly useful result from the libraries
is the Krein-Milman-Minkowski theorem,
which states that a compact convex set is the
convex hull of its extreme points \cite{Voigt2020}.

\paragraph{Polygon Definition.}
The existing library did not cover the definition of a polygon,
which we develop.
Our formal definition of a polygon mirrors Harrison's \cite{DBLP:journals/mscs/Harrison11}.
We introduce the \mpp\ function to build a polygonal
path given a list of vertices.
This function takes as input a list of points
(type \isa{(real\isacharcircum 2)\ list}) and returns a function
(of type \isa{real}\ \isasymRightarrow\ \isa{real\isacharcircum 2})
which continuously maps the interval $[0, 1]$ to $\mathbb{R}^2$.
More precisely, in Isabelle, we write the following:

\begin{isabelle}
    \mppdef
\end{isabelle}

Here, on an empty list, \mpp\ returns a path, \isa{linepath\ 0\ 0}, whose range is the origin (as a default value).
On a singleton list \isa{[a]}, it returns a path whose range is the point \isa{a}.
On a list with two points, it returns the line between those two points.
On a list with more than two points, it recursively uses the existing library function \isa{+++} to join together paths, ultimately returning a path that passes through those points in the order they are input.\footnote{Note that the \isa{+++} function introduces a particular parameterization of a path; we comment on this parameterization and challenges it introduces in \rref{sec:PolygonProperties}.}

We then define the predicate \isa{polygon} on paths,
which (intuitively) holds for a path \isa{g} iff \isa{g} is a polygon.
Our formal definition in Isabelle is as follows,
where \isa{polygonal\_path\ g} holds iff \isa{g} is in the range of \mpp,
\isa{simple\_path~g} holds iff \isa{g} is a simple path,
and \isa{closed\_path\ g} holds iff \isa{g} is a closed path
(i.e., it starts and ends at the same point).

\begin{isabelle}
\polygondef
\end{isabelle}

\paragraph{The Frontier, Interior, and Inside.}
We make frequent use of the (standard mathematical notions of)
\tit{interior} and \tit{frontier} (boundary)
of a subset $S$ of a topological space
(in our setting, $\R^2$).
We also make use of the \tit{relative interior};
the relative interior of a set $S$ is the set of all $x \in S$ such that there exists
an open set $U$ such that $x \in U \cap \text{aff}(S) \subseteq S$,
where $\text{aff(S)}$ is the affine hull of $S$.
In our formalization,
we mainly care about the relative interior of the image of a linepath,
which is simply the image of the linepath minus the endpoints. %
All of these concepts are already formalized in Isabelle's standard libraries.

Pick's theorem is a statement about the area of the ``inside'' of a polygon.
To establish this notion of inside,
we use the Jordan curve theorem (and a variant thereof) from the libraries,
which proves that every simple closed curve has an inside and an outside that satisfy some intuitively obvious properties:
1) they are disjoint from each other and from the image of the curve,
2) any point in the plane is either inside the curve, outside the curve, or on the curve,
and 3) the image of the curve is the frontier of both its inside and its outside.

\begin{figure}
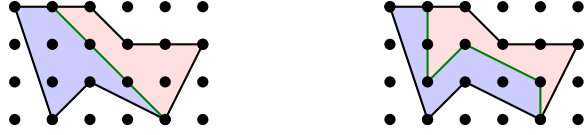

    \centering
    \begin{subfigure}{0.4\textwidth}
        \centering
        \goodlinepath{0.5}
    \end{subfigure}
    \begin{subfigure}{0.4\textwidth}
        \centering
        \goodpolygonalpath{0.5}
    \end{subfigure}
    \caption{Splitting a polygon with a good linepath and a good polygonal path.}
    \label{fig:splitting}
\end{figure}

\paragraph{Splitting a Polygon.}
Our proof of Pick's theorem relies heavily on (strong) induction
and involves splitting a polygon into two smaller polygons.
We define a \textit{good linepath} of a polygon $p$
to be a line between two different vertices of $p$ which lies entirely inside $p$,
except for its endpoints (which are on the frontier of $p$).
We generalize this notion to a \textit{good polygonal path} of $p$,
where we replace the linepath with a simple polygonal path.
In \rref{fig:splitting},
we illustrate how a good linepath and a good polygonal path split a polygon into two smaller pieces.

Our definitions of good linepath and good polygonal path
are designed to satisfy the hypotheses of the Jordan triple curve theorem, a variant of the Jordan curve theorem
which has been formalized in the libraries.
Intuitively,
the Jordan triple curve theorem makes rigorous the notion of splitting a shape into
two disjoint shapes whose union is the original shape.

We set up a lemma called \isa{pick\_union},
which states that if a polygon $p$ splits into polygons $q_1$ and $q_2$
(either by a good linepath or a good polygonal path),
and Pick's theorem holds for any two of $p$, $q_1$ and $q_2$,
then Pick's theorem holds for the remaining polygon.
From the Jordan triple curve theorem,
we easily have that the areas of the insides of $q_1$ and $q_2$
sum to the area of $p$,
and we only have to account for the vertices on the splitting path.

\subsection{The Triangle Case}\label{sec:approach-base-case}
To formalize the base case of our induction,
which amounts to formalizing Pick's theorem for triangles,
we follow steps similar to Harrison \cite{DBLP:journals/mscs/Harrison11}
and induct on $I + B$, where $I$ is the number of integral lattice points
inside $p$, and $B$ is the number of integral lattice points on
the boundary of $p$.

The base case is where $I + B = 3$,
which in particular means $I = 0$ and $B = 3$.
We prove that every such ``elementary triangle'' has area $1/2$ in a similar fashion to Harrison:
we first show that any linear transformation $L \from \R^2 \to \R^2$
such that $L(\Z^2) = \Z^2$ has determinant $\pm 1$,
then show that every elementary triangle is the image of the \textit{unit triangle} under one of these
linear transformations (modulo translation),
where the unit triangle is the convex hull of $\{(0, 0), (0, 1), (1, 0)\}$.
As the unit triangle has area $1/2$, so does every elementary triangle,
since linear transformations with determinant $\pm 1$ preserve area.

\begin{figure}
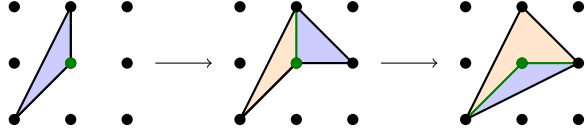

    \centering
    \trianglethreesplit{0.75}
    \caption{Applying \isa{pick\_union} twice.}
    \label{fig:trianglethreesplit}
\end{figure}

In the inductive step,
we have $I + B \geq 4$.
Then either $I \geq 1$ or $B \geq 4$.
If $I \geq 1$,
we split the triangle into three smaller triangles;
otherwise, if $B \geq 4$, we split the triangle into two smaller triangles.
In the case where we split into two triangles,
we show that the splitting linepath is a good linepath and apply \isa{pick\_union}.
In the case where we split into three triangles,
we apply \isa{pick\_union} twice
to reconstruct the inductive result on the three smaller triangles
into the result for the bigger triangle.
The second application of \isa{pick\_union} requires ``adjoining'' two polygons
whose boundary intersection is not just a linepath,
but a polygonal path of two linepaths (\rref{fig:trianglethreesplit}).
We make use of our generalization of \isa{pick\_union} to polygonal path splits to prove this.

\subsection{The Convex Case}\label{sec:approach-convex}
In the case when $p$ is convex (and not a triangle),
we find a good linepath with which we can split $p$ into
two smaller polygons to apply our inductive hypothesis.
Mathematically,
it is obvious that a convex polygon with more than three vertices has a good linepath---the line between any two non-adjacent vertices is a good linepath.
Formally, however,
this fact is not immediate,
and we outline our approach here.

Let $A$ be the convex hull of the path image of $p$.
Since $p$ is a convex polygon,
its inside is the interior of $A$,
and its path image is the frontier of $A$.
To find a good linepath,
we first apply the fact that if a linepath $\ell$ between two vertices of $p$ has a
non-empty intersection with the
interior of $A$,
then $\ell$ is in fact a good linepath.
This is a consequence of the following general property of subsets of convex sets:
if $B$ is convex and $A \subseteq B$,
then the relative interior of $A$ is a subset of the
relative interior of $B$.
This reduces the problem to simply finding a linepath which
intersects the interior of $A$. To do this,
we case on the number of extreme points $E$ of the convex hull.
As $p$ is a polygon,
we have $E \geq 3$.

\begin{figure}
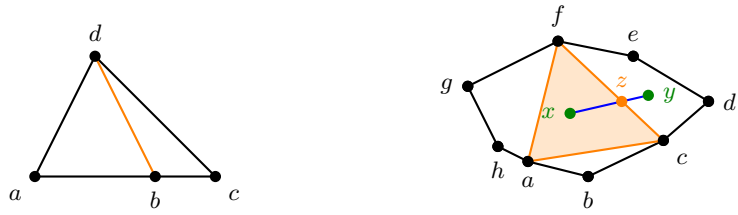

    \begin{subfigure}{0.5\textwidth}
        \centering
        \goodlinepathconvextriangle{0.8}
    \end{subfigure}
    \begin{subfigure}{0.5\textwidth}
        \centering
        \goodlinepathconvex{0.8}
    \end{subfigure}
    \caption{Finding a good linepath in a convex polygon with more than three vertices, when the convex hull has three extreme points (left) and more than three extreme points (right).
    Note that even though the polygon on the left is geometrically a triangle, it is formally a quadrilateral,
    as it has four vertices ($a$, $b$, $c$, and $d$). }
    \label{fig:convexcase}
\end{figure}

If $E = 3$,
then there is a vertex $d$ of $p$ which is not an extreme point
of $A$ (as we are not in the base case, $p$ has at least four vertices).
We then identify the extreme point of $A$ which is not on the same linepath as $d$,
and take the linepath between $d$ and this point;
this linepath is a good linepath (illustrated in \rref{fig:convexcase}).

If $E > 3$,
we take three distinct extreme points of $A$,
call them $a$, $b$, $c$,
and show that one of the linepaths of $T = \mpp\ [a, b, c, a]$
is a good linepath.
For this, it is enough to show that $T$ intersects the interior of $A$.
The argument is as follows.
Since we are in the $E > 3$ case,
the inside of $T$ is a strict subset of the interior of $A$.
So,
we can obtain points $x$ inside $T$
and $y$ in the interior of $A$ but outside $T$.
As the linepath from $x$ to $y$ intersects both the inside and outside of $T$,
it must intersect $T$ at some point $z$ (see \rref{fig:convexcase});
this is a simple corollary of a lemma from the formalization of the Poincar\'{e}-Bendixson theorem \cite{DBLP:conf/cpp/ImmlerT20}.
Since both $x$ and $y$ are in the interior of $A$,
and the interior of $A$ is convex,
we conclude that $z$ is in the interior of $A$,
as desired.

Having shown that $p$ has a good linepath,
we use it to split $p$ and apply our inductive hypothesis to the two resulting polygons (this works because each of our smaller
polygons has fewer vertices than our original polygon by construction),
then apply \isa{pick\_union}. This proves Pick's theorem for convex polygons.

\subsection{The Non-Convex Case}\label{sec:approach-non-convex}
We start our discussion of the non-convex case by giving some high-level geometric intuition, and then discuss the details of formalizing this intuition.
When $p$ is not convex,
we let $A$ be the convex hull of the path image of $p$,
and use $A$ to find a linepath $\ell$ between two vertices of $p$
that is \textit{fully outside} $p$.
In particular, we construct a sublist
$\isa{pocket\_path\_vts} = [a, x_1, \dots, x_m, b]$ of the vertices of $p$,
where $a$ and $b$ are on the frontier of $A$
and each $x_i$ is in the interior of $A$.
To simplify our construction, we assume WLOG
that this sublist begins at the first vertex of $p$.%
\footnote{Mathematically,
this WLOG assumption is immediate;
formally,
this requires rotating the vertices of a polygon, which changes its parametrization.
Showing that the new parametrization is a simple path is involved; we discuss this in \rref{sec:rotation}.}
Then we let $\ell$ be the linepath from $b$ to $a$.

\begin{figure}
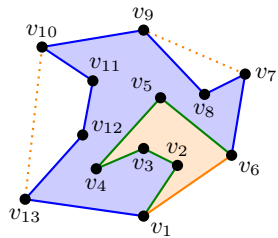

    \centering
    \pocket{0.9}
    \caption{The blue polygon is $p$,
    the orange polygon is a pocket,
    and the orange linepath from $v_1$ to $v_6$ is $\ell$, the ``filling linepath''.
    The vertex list $[v_1, v_6, v_7, \dots, v_{13}, v_1]$ generates the filled polygon.
    Dotted orange lines visualize the other pockets.}
    \label{fig:pockets}
\end{figure}

Drawing $\ell$ creates a ``filled'' polygon which is the union of
our original polygon and a missing piece of the convex hull,
which we call a \textit{pocket}.
This pocket is itself a polygon,
formed by the vertex list $[a, x_1, \dots, x_m, b, a]$. %
Ultimately, then, the construction produces two polygons that satisfy our inductive hypothesis:
the pocket and the ``filled'' polygon.
Moreover,
\isa{pocket\_path\_vts} forms a good polygonal path which splits the filled polygon
into the pocket and the original polygon, which we use to apply \isa{pick\_union}.
\rref{fig:pockets} visualizes an example.

To formalize this argument,
we first show that the filled shape,
which we call \isa{filled},
is a polygon;
as a corollary of this we obtain that the pocket,
which we call \isa{pocket},
is a polygon.
Second,
we show that \isa{pocket\_path\_vts} is a good polygonal path
(and that it splits \isa{filled} into $p$ and \isa{pocket}). %

\subsubsection{Showing that \isa{filled} is a polygon.} %
\begin{figure}
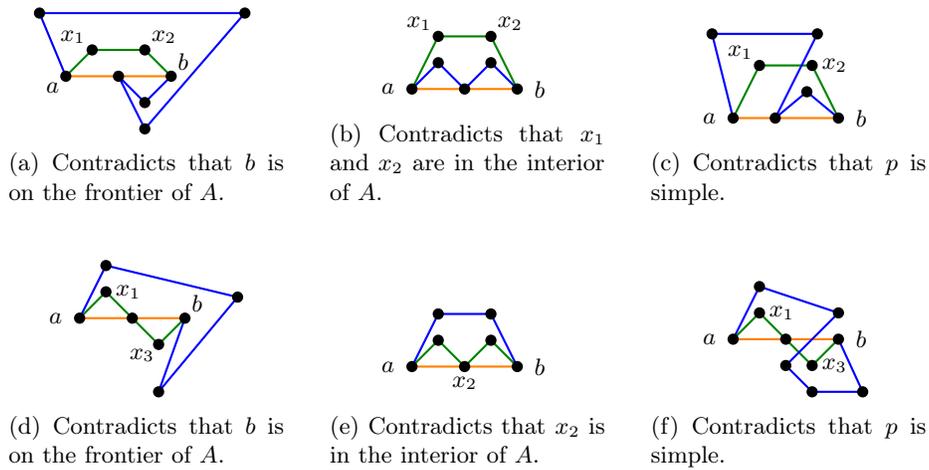

    \centering
    \begin{subfigure}{0.3\textwidth}
        \centering
        \badfillinglinepathA{1.4}
        \caption{Contradicts that $b$ is on the frontier of $A$.}
        \label{fig:filling-linepath-bad-examples-a}
    \end{subfigure}
    \hfill
    \begin{subfigure}{0.3\textwidth}
        \centering
        \badfillinglinepathB{1.4}
        \caption{Contradicts that $x_1$ and $x_2$ are in the interior of $A$.}
        \label{fig:filling-linepath-bad-examples-b}
    \end{subfigure}
    \hfill
    \begin{subfigure}{0.3\textwidth}
        \centering
        \badfillinglinepathC{1.4}
        \caption{Contradicts that $p$ is simple.}
        \label{fig:filling-linepath-bad-examples-c}
    \end{subfigure}

    \vspace{2em}
    \begin{subfigure}{0.3\textwidth}
        \centering
        \badfillinglinepathD{1.4}
        \caption{Contradicts that $b$ is on the frontier of $A$.}
        \label{fig:filling-linepath-bad-examples-d}
    \end{subfigure}
    \hfill
    \begin{subfigure}{0.3\textwidth}
        \centering
        \badfillinglinepathE{1.4}
        \caption{Contradicts that $x_2$ is in the interior of $A$.}
        \label{fig:filling-linepath-bad-examples-e}
    \end{subfigure}
    \hfill
    \begin{subfigure}{0.3\textwidth}
        \centering
        \badfillinglinepathF{1.4}
        \caption{Contradicts that $p$ is simple.}
        \label{fig:filling-linepath-bad-examples-f}
    \end{subfigure}
    \caption{Examples of $\ell$ intersecting $p$ at a point other than an endpoint of $\ell$.
    }
    \label{fig:filling-linepath-bad-examples}
\end{figure}

When determining how to formalize that $\ell$ intersects $p$ only at $a$ and $b$
(which is sufficient to show that \isa{filled} is a polygon),
we found it useful to draw various examples
where this does not hold,
to understand how each violates the properties of our construction;
\rref{fig:filling-linepath-bad-examples} shows some examples.
We orient the figures so that the ``filling linepath'' $\ell$ is horizontal,
as our formalization assumes WLOG that $\ell$ lies
on the $x$-axis.%

\newcommand{\caseA}{\rref{fig:filling-linepath-bad-examples-a}\xspace}
\newcommand{\caseB}{\rref{fig:filling-linepath-bad-examples-b}\xspace}
\newcommand{\caseC}{\rref{fig:filling-linepath-bad-examples-c}\xspace}
\newcommand{\caseD}{\rref{fig:filling-linepath-bad-examples-d}\xspace}
\newcommand{\caseE}{\rref{fig:filling-linepath-bad-examples-e}\xspace}
\newcommand{\caseF}{\rref{fig:filling-linepath-bad-examples-f}\xspace}

Organizing the various possible contradictory situations illustrated
in \rref{fig:filling-linepath-bad-examples}
into a collection of formal lemmas was challenging.
We first preclude the \caseB and \caseE cases
by showing that no $x_i$ can have the smallest or largest $y$-coordinate.
We then show that no point on $p$ can have negative $y$-coordinate,
which rules out the cases in \caseA, \caseD, and \caseF.
Finally, in the \caseC case,
we show that $p$ is not simple,
a contradiction.

To show that no $x_i$ can have the smallest or largest $y$-coordinate,
we simply show that any point on $p$ which has smallest or largest $y$-coordinate
is on the frontier of $A$.
This precludes the cases in \caseB and \caseE.
Moreover,
we obtain some vertex $y_r$ which has smaller $y$-coordinate than every $x_i$,
and some vertex $y_s$ which has larger $y$-coordinate than every $x_i$,
which we will use to rule out the other cases.

To show that no point on $p$ can have negative $y$-coordinate,
we assume for contradiction that $y_r$ has negative $y$-coordinate.
Assuming WLOG that $r < s$,
we consider where the subpath of $p$ from $y_r$ to $y_s$ intersects the $x$-axis.
We first prove that neither \isa{pocket\_path} nor the subpath can intersect the $x$-axis at a point
not lying on $\ell$,
as such an intersection would put $a$ or $b$ in the interior of $A$.
This precludes the cases \caseA and \caseD,
leaving us with the cases in \caseC and \caseF.

\begin{figure}
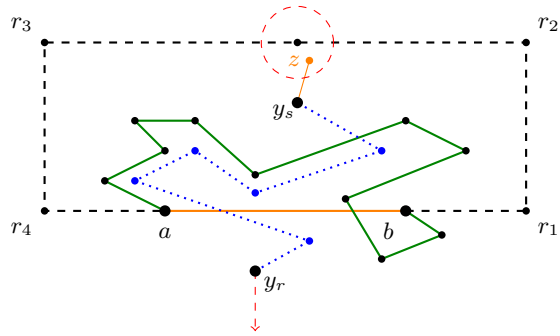

    \centering
    \subpathintersectionA{1.6}
    \caption{If the subpath and \isa{pocket\_path} intersect the $x$-axis only on $\ell$,
    then the subpath intersects \isa{pocket\_path}.}
    \label{fig:subpath-intersection}
\end{figure}

In the \caseF case,
the subpath and \isa{pocket\_path} intersect the $x$-axis only on $\ell$,
and here we show that the subpath must intersect \isa{pocket\_path}
(which contradicts that $p$ is simple).
For this, we construct an axis-parallel rectangle whose lower edge lies on the $x$-axis,
and whose width and height are large enough so that the entire pocket path and subpath
are inside the rectangle.
Then,
we delete the points lying between $a$ and $b$ on the bottom edge of the rectangle
and replace the resulting gap by \isa{pocket\_path} (illustrated in \rref{fig:subpath-intersection});
we let $R$ be the resulting closed, simple path.
We show that $y_r$ is outside $R$,
and that $y_s$ is inside $R$,
and this yields (via a lemma from the Poincar\'{e}-Bendixson formalization \cite{DBLP:conf/cpp/ImmlerT20})
that the subpath intersects $R$.
However,
by the construction of $R$,
the subpath can neither intersect the side edges nor the top edge of $R$,
nor can it intersect the two axis-parallel bottom edges of $R$.
Thus,
the only remaining part of $R$ which it can intersect is \isa{pocket\_path},
as desired.

Formalizing that $y_r$ is outside $R$
and $y_s$ is inside $R$ in the above argument is itself a challenge.
The construction of $R$ makes this pictorially obvious, but formally,
we only know the abstract facts about the inside and outside of $R$
given by the Jordan curve theorem (see \rref{sec:prelim}).
Working with the abstract definitions, we show that $y_r$
is in the unbounded component of $\R^2 \setminus R$ (i.e., $y_r$ is outside $R$)
by constructing an infinite ray beginning at $y_r$ which never intersects $R$
(a downward vertical ray suffices).
To obtain that $y_s$ is inside $R$,
we take a sufficiently small $\epsilon$-ball around the point on $R$
directly above $y_s$,
and (using a library result) obtain a point $z$ which is both in this $\epsilon$-ball and inside $R$.
We then show that $z$ cannot be above the upper edge of $R$
(if it were, it would be outside $R$, which we show by taking an upward vertical ray).
Finally,
we take the linepath from $z$ to $y_s$ and show that this linepath cannot intersect $R$;
then, as $z$ is inside $R$, so is $y_s$.
These proof techniques are illustrated in \rref{fig:subpath-intersection}.

So, $y_r$ cannot have negative $y$-coordinate,
which establishes that every point on $p$ has non-negative $y$-coordinate.
Then, since $\ell$ lies on the $x$-axis, $\ell$ lies on the frontier of $A$.
Thus,
\isa{pocket\_path} does not intersect $\ell$ at a point other than its endpoints,
precluding the \caseF case.%
\footnote{This implies that \isa{pocket} is a polygon,
which is used later in the proof that \isa{pocket\_path} is a good polygonal path.}

The case in \caseC is the only remaining case.
We assume for contradiction that $\ell$ minus its endpoints intersects $p$,
and with a bit of work, we can assume in particular that $y_r$ lies on $\ell$.
So, we have a subpath (from $y_r$ to $y_s$) of $p$ which starts at a point on $\ell$
and ends at a point above all points on \isa{pocket\_path},
illustrated in \rref{fig:subpath-intersection-2}. 
We can then show that this subpath necessarily crosses \isa{pocket\_path}
using some of our prior proof techniques
(compare \rref{fig:subpath-intersection}).

\begin{figure}
    \centering
    \subpathintersectionB{1.6}
    \caption{If all points on $p$ are non-negative, we have a subpath of $p$ which starts on $\ell$ and ends at a point above all points of $p$, and thus $p$ intersects itself.}
    \label{fig:subpath-intersection-2}
\end{figure}

\subsubsection{Showing that \isa{pocket\_path} is a good polygonal path of \isa{filled}.}
From the previous result, we have that \isa{pocket\_path} intersects \isa{filled} only at $a$ and $b$,
and so \isa{pocket\_path} (minus $a$ and $b$)
is either entirely inside or entirely outside \isa{filled}.\footnote{As we now have that \isa{filled} is a polygon,
we can refer to its inside and outside.}
Thus,
to show that \isa{pocket\_path} is inside \isa{filled},
it suffices to find only a single point on \isa{pocket\_path} which is inside \isa{filled}.

\begin{figure}
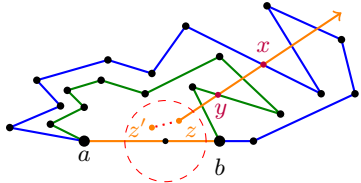

    \centering
    \pocketpathgood{0.9}
    \caption{The ray $r$, shown in orange,
    intersects \isa{pocket\_path} at $y$
    before its first intersection with \isa{filled} at $x$.}
    \label{fig:pocket-path-good}
\end{figure}

We first obtain a point $z$ which is inside both \isa{pocket} and \isa{filled}.
While it is intuitively clear that the inside of \isa{pocket}
is a subset of the inside of \isa{filled},
showing this formally requires some machinery.
We take an $\epsilon$-ball around a point on $\ell$,
and obtain points $z$ and $z'$ in the $\epsilon$-ball
where $z$ is inside \isa{pocket} and $z'$ is inside \isa{filled}; this is illustrated in \rref{fig:pocket-path-good}.
We ensure that $\epsilon$ is small enough so that the linepath from $z$ to $z'$
does not intersect \isa{filled},
and this shows that $z$ is inside \isa{filled}.

So, we have that $z$ is inside both \isa{pocket} and \isa{filled},
and we take a ray $r$ starting at $z$,
constructed so that $r$ intersects \isa{pocket\_path} at a point $y$
before its first intersection with \isa{filled} at a point $x$.
As $r$ starts inside \isa{filled},
all points on $r$ prior to $x$ are inside \isa{filled},
and thus $y$ is inside \isa{filled} and on \isa{pocket\_path}.
As we know that \isa{pocket\_path} is simple and intersects \isa{filled} only at $a$ and $b$,
the existence of $y$ suffices to show that \isa{pocket\_path} is~a~good~polygonal~path~of~\isa{filled}.

\subsection{Top-Level Result}\label{sec:toplevel}
Putting all the pieces together yields the following top-level result.
\begin{isabelle}
    \pick
\end{isabelle}
This theorem states that if \isa{p} is a polygon with vertex list \isa{vts}
where all of the vertices in \isa{vts} are integral lattice points (assumption \isa{all\_integral vts}),
and if \isa{I} is the cardinality of the set of integral points inside \isa{p}
and \isa{B} is the cardinality of the set of integral lattice points on the boundary of \isa{p},
then the area of \isa{p} (given by the Lebesgue measure of its path inside,
using Isabelle/HOL's standard library definitions \isa{measure} and \isa{lebesgue})
is \isa{I + B/2 - 1}, as claimed by Pick's theorem.

\section{Formalization Details}\label{sec:formalization}
Having outlined our proof approach, we turn to some further details of our formalization,
including challenges we faced and library extensions we contribute.

\subsection{Polygon Properties}\label{sec:PolygonProperties}
\subsubsection{Linepath Characterization of Polygonal Paths.}
\begin{figure}
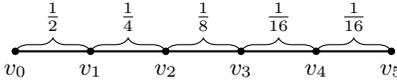

    \centering
    \polygonalpathparametrization{1}
    \caption{Paramaterization of a polygonal path with six vertices.}
    \label{fig:parametrization}
\end{figure}

In the Isabelle libraries,
a path is a map from the unit interval $[0, 1]$
into some topological space.
The library definition of joining two paths $p_1$ and $p_2$ together, \isa{+++},
assigns the first half of the unit interval to $p_1$,
and the second half to $p_2$.
We construct polygons by repeated application of this \isa{+++} operation (see \rref{sec:prelim}).
This has the effect that the first linepath in a polygonal path corresponds to
the interval $[0, \frac{1}{2}]$,
the second linepath to $[\frac{1}{2}, \frac{3}{4}]$,
the third linepath to $[\frac{3}{4}, \frac{7}{8}]$,
and so forth;
the last linepath corresponds to $[\frac{2^{n-1} - 1}{2^{n-1}}, 1]$,
where $n$ is the number of vertices in the polygonal path (see \rref{fig:parametrization}).
Harrison also used this parametrization in his work \cite{DBLP:journals/mscs/Harrison11}.
Using \isa{+++} in this way allows us to directly apply library lemmas about joining paths,
but we found this parametrization unwieldy to work with in practice.

For example,
suppose $p = \mpp\ \isa{[a, b, c, d, a]}$
is a polygon.
Then we might expect
\begin{equation*}
    p' = \isa{(make\_polygonal\_path [a, b, c]) +++ (make\_polygonal\_path [c, d, a])}
\end{equation*}
to also form a polygon.
While we can show that $p'$ is simple and has the same path image as $p$,
$p'$ is (unintuitively) not a polygon because of its parametrization.
We also encounter challenges when proving that
if \isa{vts} generates a simple polygonal path $p$,
then so does any sublist of length at least two of \isa{vts}.
While library lemmas establish that a subpath of a simple path is simple,
our polygon parametrization means that
a sublist does not necessarily generate a \tit{subpath}.

To mitigate such challenges,
we ultimately formalized a collection of lemmas that translate between the
parametrization of a polygonal path $p$
and the parametrization of its constituent linepaths:
given $t \in [0, 1]$,
we obtain $t' \in [0, 1]$ where $p(t) = \ell(t')$
and $\ell$ is a linepath of $p$.
Though not overly mathematically complex,
this translation was somewhat tedious to formalize and apply.

\subsubsection{Polygon Vertex Rotation.}\label{sec:rotation}
We often found it convenient to assume WLOG that a property which holds for \textit{some}
vertex in a polygon holds for the \textit{first} vertex.
Intuitively, if  $p = \mpp\ \isa{[a, b, c, d, a]}$
is a polygon, then it is ``essentially'' the same polygon as
$p' = \mpp\ \isa{[b, c, d, a, b]}$.
Though $p$ and $p'$ are different curves, they have the same path image and are geometrically the same.
However,
it was non-trivial to formalize that $p$ is a polygon iff $p'$ is a polygon because they are parametrized differently (this was our original impetus to formalize the aforementioned parametrization translation between a polygonal path and its linepaths).

In Isabelle,
we formalize rotation on the vertex list of a polygon in the function \isa{rotate\_polygon\_vertices}
and prove the following crucial properties to establish that
rotating a polygon yields a polygon with the same path image.
\begin{isabelle}
\rotationlemma
\end{isabelle}

\subsection{Unit Geometry}
In the base case of our induction to prove Pick's theorem for triangles,
we must show that the area of the unit triangle is $1/2$ (see \rref{sec:approach-base-case}).
While the formula for the area of a triangle is a grade-school fact,
we formally treat the area of a polygon as the Lebesgue measure of its inside,
which is not entirely straightforward to work with (given the abstractness
of the Lebesgue measure and `inside' definitions).
We use existing library results to show that the area of a unit square is $1$,%
\footnote{Intuitively,
the Lebesgue measure of open rectangles is ``hard-coded'' in the definition;
showing that the area of the unit square is $1$ essentially amounts to showing
that the unit square (defined as a convex hull of its four corners)
is $[0, 1] \times [0, 1]$.}
and then show that the unit square can be split into the unit triangle and its diagonal mirror.
As the unit triangle and its mirror have the same area, and
their areas sum to $1$,
they each must have area $1/2$.
Formalizing this involved concretely characterizing the abstract `inside'
of the unit triangle and unit square (for example, we prove that the unit triangle is the set of points $(x, y)$ where $0 \leq x$, $0 \leq y$, and $x + y \leq 1$) and matching between the various definitions (e.g., the interior of the convex hull of three points is the same as the path inside of the polygon formed by the three points),
a task which overall proves more intricate and finicky than
grade-school intuition would suggest.

\subsection{Convex Hull Properties}
In the course of our formalization, we need to prove a variety of general results relating
convex hulls to polygons.
This can be tricky,
since a polygon is a mapping with specific properties,
whereas a convex hull is simply a set of points.
For example, if we take the convex hull of a polygon
(which is the same as taking the convex hull of the vertices of the polygon),
then \textit{geometrically} the result can clearly be characterized as a polygon.
In generality, the missing result is that the convex hull of any finite set of points can be alternatively characterized by the path inside of some polygon.
Formally establishing this connection is nontrivial; in our proof approach, we formalize a significant step towards this by showing that filling in a pocket of the convex hull yields a polygon.

We additionally contribute other results that augment the existing library support for convex hulls.
For example, we prove that if all of the vertices of a polygon
are on the frontier of its convex hull,
then the polygon itself is convex.
In contrapositive, this yields that for a non-convex polygon,
some vertex is in the interior of its convex hull,
which we use to show that a non-convex polygon has a pocket.
While geometrically obvious in a proof-by-picture fashion,
the formal proof was more detailed and involved relating the notions of convex hulls,
affine hulls, and half planes;
in particular, we show that if a linepath is not on the frontier
of the convex hull, it ``cuts'' the convex hull,
which then forces a self-intersection in the polygonal path.
This notion of ``cutting'' is different than that afforded by the Jordan triple curve theorem,
and its formalization involved
characterizing certain intersections of a convex hull with an affine hull of two points,
as well as classifying the frontier of half plane as an affine hull.

\section{Conclusion and Future Work}\label{sec:conclusion}
Our formalization of Pick's theorem,
the second in any theorem prover,
involves both creative strategies to formalize pictorial arguments
and considerable fundamental library extensions.
The other existing formalization of Pick's theorem,
by John Harrison in HOL Light \cite{DBLP:journals/mscs/Harrison11},
emphasizes the surprisingly involved nature of formalizing geometry.
While we succeeded in avoiding a particularly thorny part of Harrison's proof
by formalizing a less common approach to proving Pick's theorem,
we do not claim that our formalization is overall simpler.
Ultimately, we resonate with many of the difficulties Harrison describes.
Describing the step to verify the existence of a good linepath in a polygon,
Harrison writes ``. . . we found it quite hard work reasoning about
obvious facts like `this point and that point are on opposite sides of the line'.''
\cite[p. 12]{DBLP:journals/mscs/Harrison11}.
We encountered a similar difficulty in our formalization,
particularly in the piece of our proof that avoided this very step
(the non-convex case, \rref{sec:approach-non-convex}).

While we believe there are inherent challenges in formalizing geometry,
a retrospective analysis also reveals portions of our formalization that
were potentially unnecessarily painful.
In particular, the difficulties we faced with the parametrization induced by
our polygon construction could have potentially (retrospectively) been avoided
by using a polygon definition along the following lines:
$P \subseteq \R^2$ is a polygon iff (1) $P$ is the \textit{path image} of some (potentially non-simple)
polygonal path,
and (2) $P$ is the \tit{path image} of some simple path.
Note that this definition treats a polygon as a subset of $\R^2$,
not a mapping from $[0, 1]$ into $\R^2$.
This definition lets us find separate witnesses of the ``polygonal-ness''
and ``simple-ness'' of a polygon,
which may avoid the problems we encountered in showing that certain
vertex transformations of polygons,
such as polygon vertex rotation or taking sublists,
yield simple polygonal paths. %

In addition to exploring potential simplifications of our formalization,
there are some interesting possible extensions of our work.
Our formalization takes the first step towards showing that
any polygon can be made convex by removing some of its vertices
(i.e., by filling in its pockets).
Further, our formal treatment of pockets opens up the possibility of
verifying other theorems about pockets,
such as the Erd\"{o}s-Nagy theorem,
which states that any polygon can be made convex by successively picking
a pocket and flipping it across
the face it shares with the convex hull \cite{grunbaum1995convexify}.
We are also interested in other theorems related to area and geometry,
e.g. the Shoelace theorem (alternately called the Surveyor's Formula) \cite{braden1986surveyor}, and in connecting our work to some of the pre-existing Isabelle results like Ceva's theorem (which used a different definition of area) \cite{Ceva-AFP}.

\begin{credits}
\subsubsection{\ackname} This material is based upon work supported by the National Science Foundation under Grant No. 1552934. 
Thank you to Lawrence and Susan Sheets for generously supporting S.B.
with a graduate fellowship.
Thank you to Aaron Stump and Yong Kiam Tan for useful comments on the paper.

\subsubsection{\discintname}
The authors have no competing interests to declare that are relevant to the content of this article.
\end{credits}
 \bibliographystyle{splncs04}
 \bibliography{pick}

\end{document}